\documentclass[12pt,index=totoc,openany]{article}
\usepackage{amsmath,amsfonts,amssymb,amsthm,a4wide,graphics}
\usepackage{makeidx,showidx}
\usepackage{nameref}
\usepackage[pdftex,bookmarks=true,bookmarksnumbered]{hyperref}
 \hypersetup{colorlinks=true,%
            citecolor=black,%
            filecolor=black,%
            linkcolor=black,%
            urlcolor=black,%
         pdfsubject={},%
        pdfnewwindow=true,%
        pdffitwindow=true,%
        pdftoolbar=true,%
        pdfmenubar=true,%
        bookmarks=true,%
        unicode=false,%
                pdftex}

 \numberwithin{equation}{section}

\begin{document}

%%%%%%%%%%%%%%%%%%%%%%%%%%%%%%%%%%%%%%%%%%%%%%%%%%%%%%%%%%%%%%%%%%

%%%%%%%%%%% dole: nlqm2abs.tex %%%%%%%%%
\def\nl{\hfil\break}
%\def\m#1{\mbox{\ensuremath{#1}}}

%%%%%%%%%%hore: eqm1-x-abs %%%%%%%%%%

\hyphenation{co-ad-joint re-pre-sen-ta-ti-on Bra-ti-sla-va}

\def\refer#1#2#3#4#5{#1,\ {\sl #2}\ {\bf #3}\ {(#4)}\ #5;\ }

\def\acc{$\!\!$\'{}$\!$}
\def\dti{\!\cdot\!}
\def\rref#1~{~(\ref{eq;#1})}
\def\dref#1~{~Definition~\ref{df;#1}}
\def\pgr#1~{~\pageref{eq;#1}}
\def\pgd#1~{~\pageref{df;#1}}
\def\pg#1~{~\pageref{#1}}
\def\bs#1{$\boldsymbol{#1}$}
\def\mbs#1{\boldsymbol{#1}}

\def\Ran{{\rm Ran}}
\def\Ker#1{${\rm Ker}(#1)$}
\def\mKer#1{{\rm Ker}(#1)}

\def\supp{{\rm supp}\ }

\def\deg#1{{\rm deg}($#1$)}
\def\mdeg#1{{\rm deg}(#1)}

\def\mT#1{\mcl T_{#1}}
\def\T#1{$\mcl T_{#1}$}
\def\cw{\curlywedge}

\def\LM#1{$\mbs{\Lambda^{#1}}(M)$}
\def\mLM#1{\mbs{\Lambda^{#1}}(M)}
\def\LN#1{$\mbs{\Lambda^{#1}}(N)$}
\def\mLN#1{\mbs{\Lambda^{#1}}(N)}

\def\mSs{{\cal S}_*}
\def\Ss{${\cal S}_*$}
\def\cS{${\cal S}$}
\def\mS{{\cal S}}
\def\Ej{$E_{\{j\}}$}
\def\mEj{E_{\{j\}}}
\def\mrj{\mrh_{\{j\}}}
\def\rj{$\mrh_{\{j\}}$}
\def\psitr{\psi_t^\mrh}

%%%Pouzivat jedine v math-mode!:
\def\mbf#1{{\mathbf #1}}
%%%%%%%%%%%%%%%%%%%%%%%%%%%%

%\def\pika{\ $\diamondsuit$\nl}
\def\pika{\ $\diamondsuit$}
\def\bpika{\ $\blacklozenge$}
\def\zel{\ $\spadesuit$}
\def\zal{\ $\clubsuit$}
\def\dovi{\ $\heartsuit$}

\def\pI{${\mbf p}_I$}
\def\pII{${\mbf p}_{II}$}
\def\mpI{{\mbf p}_I}
\def\mpII{{\mbf p}_{II}}

\def\bT{$\Bbb T$}
\def\mbT{{\Bbb T}}
\def\bA{$\Bbb A$}
\def\mbA{{\Bbb A}}
\def\bK{$\Bbb K$}
\def\mbK{{\Bbb K}}
\def\bC{$\Bbb C$}
\def\mbC{{\Bbb C}}
\def\bR{$\Bbb R$}
\def\mbR{{\Bbb R}}
\def\bN{$\Bbb N$}
\def\mbN{{\Bbb N}}
\def\bZ{$\Bbb Z$}
\def\mbZ{{\Bbb Z}}
\def\bI{$\Bbb I$}
\def\mbI{{\Bbb I}}
\def\bF{$\Bbb F$}
\def\bG{$\Bbb G$}
\def\mbF{{\Bbb F}}
\def\mbG{{\Bbb G}}
\def\bbf{${\bf f}$}
\def\mbbf{{\bf f}}
\def\cbF{${\bf F}$}
\def\mcbF{{\bf F}}
\def\bD{{\bf D}}
\def\bx{{\bf x}}
\def\by{{\bf y}}
\def\bz{{\bf z}}
\def\bv{{\bf v}}
\def\bw{{\bf w}}

\def\bbs#1{$\boldsymbol{#1}$}
\def\mbbs#1{\boldsymbol{#1}}
\def\bbr#1{$\boldsymbol{(#1)}$}
\def\mbbr#1{\boldsymbol{(#1)}}

\def\ia{{\it a}}
\def\ib{{\it b}}
\def\ix{{\it x}}
\def\iy{{\it y}}
\def\iz{{\it z}}
\def\iu{{\it u}}

\def\fpsxe{$\rf^\psi_{[\xi,\eta]}$}
\def\mfpsxe{\rf^\psi_{[\xi,\eta]}}
\def\fpsx{$\rf^\psi_\xi$}
\def\mfpsx{\rf^\psi_\xi}
\def\fpse{$\rf^\psi_\eta$}
\def\mfpse{\rf^\psi_\eta}

\def\rq{{\rm q}}
\def\rf{{\rm f}}
\def\rrh{{\rm h}}
\def\rx{{\rm x}}
\def\ry{{\rm y}}
\def\rz{{\rm z}}
\def\ra{{\rm a}}
\def\rrb#1{{\rm b}_{#1}}
\def\rc{{\rm c}}
\def\rd{{\rm d}}
\def\rQ{{\rm Q}}
\def\ru{{\rm u}}
\def\rv{{\rm v}}
\def\mrho{{\rm h}_0}
\def\mrhoo{{\rm h}^0}

\def\cT{$\cal{T}$}

\def\CH{${{\frak C}({\cal H})}$}
\def\mCH{{{\frak C}({\cal H})}}
\def\TH{${{\frak T}({\cal H})}$}
\def\UH{${\cal U(H)}$}
\def\mTH{{{\frak T}({\cal H})}}
\def\mUH{{\cal U(H)}}
\def\Ur{${\mfk U}_\rho$}
\def\mUr{{\mfk U}_\rho}

\def\Lqn{$L^2(\mbR^n)$}
\def\mLqn{L^2(\mbR^n)}
\def\LHs{${\cal L(H)}_s$}
\def\mLHs{{\cal L(H)}_s}
\def\mH{{\cal H}}
\def\mK{{\cal K}}
\def\mLH{{\cal L(H)}}
\def\LH{${\cal L(H)}$}
\def\LHo{${\cal L(H}_\mome)$}
\def\mLHo{{\cal L(H}_\mome)}
\def\H{${\cal H}$}
\def\K{${\cal K}$}
\def\PH{$P({\cal H})$}
\def\mPH{P({\cal H})}
\def\A{${\cal A}$}
\def\mA{{\cal A}}
\def\Z{${\cal Z}$}
\def\mZ{{\cal Z}}
\def\C{${\cal C}$}
\def\Cc{${\cal C}_{cl}$}
\def\mCc{{\cal C}_{cl}}
\def\mC{{\cal C}}
\def\B{${\cal B}$}
\def\mB{{\cal B}}
\def\F{${\cal F}$}
\def\mF{{\cal F}}
\def\Fr{$\mF_{\mrh}$}
\def\mFr{\mF_{\mrh}}
\def\mFP{\mF_{\mPH}}
\def\FP{$\mF_{\mPH}$}

\def\PM{$\mcl P(\mfk M)$\ }
\def\mPM{\mcl P(\mfk M)}

\def\cl#1{${\cal #1}$}
\def\mcl#1{{\cal #1}}

\def\ben{$\beta_{\nu}$}
\def\mben{\beta_{\nu}}
\def\ber{$\beta_{\mrh}$}
\def\mber{\beta_{\mrh}}
\def\qr{$\rq_{\mrh}$}
\def\mqr{\rq_{\mrh}}
\def\pr{${\rm p}_{\mrh}$}
\def\mpr{{\rm p}_{\mrh}}
\def\qn{$\rq_{\nu}$}
\def\mqn{\rq_{\nu}}
\def\Tr{$T_{\mrh}$}
\def\mTr{T_{\mrh}}
\def\Trs{$T_{\mrh}^*$}
\def\mTrs{T_{\mrh}^*}
\def\Tn{$T_{\nu}$}
\def\mTn{T_{\nu}}
\def\Tns{$T_{\nu}^*$}
\def\mTns{T_{\nu}^*}

\def\Ty{$T_{\by}$}
\def\mTy{T_{\by}}

\def\P#1{$P_{#1}$}
\def\mP#1{P_{#1}}

\def\bcDp{${\boldsymbol{\mcl D}}_{r+}^1$}
\def\mbcDp{{\boldsymbol{\mcl D}}_{r+}^1}
\def\bcD{${\boldsymbol{\mcl D}}_r$}
\def\mbcD{{\boldsymbol{\mcl D}}_r}
\def\bcDF{${\boldsymbol{\mcl D}}(\mbF)$}
\def\mbcDF{{\boldsymbol{\mcl D}}(\mbF)}
\def\bcDFr{${\boldsymbol{\mcl D}_r}(\mbF)$}
\def\mbcDFr{{\boldsymbol{\mcl D}_r}(\mbF)}
\def\DdYa{$\mcl D_{ra}(\delta_Y)$}
\def\mDdYa{\mcl D_{ra}(\delta_Y)}
\def\DdX{$\mcl D_r(\delta_X)$}
\def\DdXa{$\mcl D_{ra}(\delta_X)$}
\def\DdXs{$\mcl D_{r*}(\delta_X)$}
\def\mDdX{\mcl D_r(\delta_X)}
\def\mDdXa{\mcl D_{ra}(\delta_X)}
\def\mDdXd{\mcl D_{rd}(\delta_X)}
\def\DdXd{$\mcl D_{rd}(\delta_X)$}
\def\mDdXs{\mcl D_{r*}(\delta_X)}
\def\DX{$\mcl D_r(X)$}
\def\DXa{$\mcl D_{ra}(X)$}
\def\DXd{$\mcl D_{rd}(X)$}
\def\DXs{$\mcl D_{r*}(X)$}
\def\mDX{\mcl D_r(X)}
\def\mDXa{\mcl D_{ra}(X)}
\def\mDXs{\mcl D_{r*}(X)}
\def\As{$A^*$}
\def\DAs{$D(A^*)$}
\def\DA{$D(A)$}
\def\HH{$\mH\oplus\mH$}
\def\mHH{\mH\oplus\mH}

\def\pEF{$\tilde{\mcl E_{\mbF}}$}
\def\mpEF{\tilde{\mcl E_{\mbF}}}
\def\EF{$\mcl E_{\mbF}$}
\def\mEF{\mcl E_{\mbF}}

\def\DGom{$\mcl D^{\mome}(G)$}
\def\mDGom{\mcl D^{\mome}(G)}
\def\DGomr{$\mcl D^{\mome}_r(G)$}
\def\mDGomr{\mcl D^{\mome}_r(G)}

\def\d#1,#2~{$d_{#2}{#1}$}
\def\md#1,#2~{d_{#2}{#1}}
\def\D#1,#2~{$D_{#2}{#1}$}
\def\mD#1,#2~{D_{#2}{#1}}
\def\Dfr{\D f,\mrh~}
\def\dfr{\d f,\mrh~}
\def\mDfr{\mD f,\mrh~}
\def\mdfr{\md f,\mrh~}
\def\Dfn{\D f,\nu~}
\def\dfn{\d f,\nu~}
\def\mDfn{\mD f,\nu~}
\def\mdfn{\md f,\nu~}
\def\Dhr{\D h,\mrh~}
\def\dhr{\d h,\mrh~}
\def\mDhr{\mD h,\mrh~}
\def\mdhr{\md h,\mrh~}
\def\mDhn{\mD h,\nu~}
\def\mdhn{\md h,\nu~}
\def\Dhn{\D h,\nu~}
\def\dhn{\d h,\nu~}

\def\mN#1,#2~{\|#1\|_{#2}}
\def\N#1,#2~{$\|#1\|_{#2}$}

\def\ad{{\rm ad}}
\def\madr{{\rm ad}_{\mrh}}
\def\madn{{\rm ad}_{\nu}}
\def\adr{${\rm ad}_{\mrh}$}
\def\adn{${\rm ad}_{\nu}$}
\def\madrs{{\rm ad}^*_{\mrh}}
\def\madns{{\rm ad}^*_{\nu}}
\def\adrs{${\rm ad}^*_{\mrh}$}
\def\adns{${\rm ad}^*_{\nu}$}

\def\OGF{$\mcl O_{F}(G)$}
\def\mOGF{\mcl O_{F}(G)}
\def\OWHr{$\mcl O_{\mrh}(\mGWH)$}
\def\mOWHr{\mcl O_{\mrh}(\mGWH)}
\def\OUr{$\mcl O_{\mrh}({\mfk U})$}
\def\OGr{$\mcl O_{\mrh}(G)$}
\def\OUn{$\mcl O_{\nu}(\mfk U)$}
\def\OGn{$\mcl O_{\nu}(G)$}
\def\OU{\cl O(\fk U)}
\def\OG{\cl O($G$)}
\def\mOUr{\mcl O_{\mrh}(\mfk U)}
\def\mOGr{\mcl O_{\mrh}(G)}
\def\mOUn{\mcl O_{\nu}(\mfk U)}
\def\mOGn{\mcl O_{\nu}(G)}
\def\mOU{\mcl O(\mfk U)}
\def\mOG{\mcl O(G)}
\def\Or{$\mcl O_{\mrh}$}
\def\mOr{\mcl O_{\mrh}}
\def\On{$\mcl O_{\nu}$}
\def\mOn{\mcl O_{\nu}}

\def\fk#1{${\frak #1}$}
\def\mfk#1{{\frak{#1}}}

\def\fTs{$\mfk T_s$}
\def\mfTs{\mfk T_s}
\def\fNn{$\mfk N_{\nu}$}
\def\mfNn{\mfk N_{\nu}}
\def\fNr{$\mfk N_{\mrh}$}
\def\mfNr{\mfk N_{\mrh}}
\def\fMr{$\mfk M_{\mrh}$}
\def\mfMr{\mfk M_{\mrh}}

\def\eps{$\epsilon$\ }
\def\meps{\epsilon}
\def\veps{$\varepsilon$\ }
\def\mveps{\varepsilon}
\def\ome{$\omega$\ }
\def\mome{\omega}
\def\gam{$\gamma$\ }
\def\mgam{\gamma}
\def\alp{$\alpha$}
\def\malp{\alpha}
\def\mphi{{\varphi}}
\def\rh{$\rho$}
\def\mrh{\rho}
\def\sg{$\sigma$}
\def\sgo{$\sigma_0$}
\def\msg{\sigma}
\def\msgo{\sigma_0}
\def\mlam{\lambda}
\def\lam{$\lambda$}

\def\Xx{$X_{\xi}$}
\def\mXx{X_{\xi}}
\def\Xe{$X_{\eta}$}
\def\mXe{X_{\eta}}
\def\Xxe{$X_{[\xi,\eta]}$}
\def\mXxe{X_{[\xi,\eta]}}

\def\Ac{$\mcl A_{cl}$}
\def\mAc{\mcl A_{cl}}
\def\Cbs{$\mcl C_{bs}$}
\def\mCbs{\mcl C_{bs}}
\def\CG{${\mcl C}^G$}
\def\mCG{{\mcl C}^G}
\def\CGc{${\mcl C}^G_{cl}$}
\def\mCGc{{\mcl C}^G_{cl}}
\def\CGq{$\mcl C^G_{q}$}
\def\mCGq{{\mcl C}^G_{q}}
\def\GGc{${\mcl G}^G_{cl}$}
\def\mGGc{{\mcl G}^G_{cl}}
\def\GG{${\mcl G}^G$}
\def\mGG{{\mcl G}^G}
\def\nGcl{$\nu G$--classical\ }
\def\rGcl{$\mrh G$--classical\ }
\def\Gcl{$G$--classical\ }

\def\GWH{$G_{WH}$}
\def\mGWH{G_{WH}}

\def\hh#1,#2,#3~{$\hat {h_{\mfk#1}}(#2,#3)$}
\def\mhh#1,#2,#3~{\hat {h_{\mfk#1}}(#2,#3)}

\def\mh#1{{h}_{#1}}
\def\h#1{${h}_{#1}$}
\def\hSl{$\mh H^{Sl}$}
\def\mhSl{\mh H^{Sl}}

\def\ph#1,#2~{$\varphi_{#1}^{#2}$}
\def\mph#1,#2~{\varphi_{#1}^{#2}}
\def\cmrh#1~{{\varrho_{#1}}}
\def\crh#1~{$\varrho_{#1}$\ }

\def\pph#1,#2~{$\tilde{\varphi}_{#1}^{#2}$}
\def\mpph#1,#2~{\tilde{\varphi}_{#1}^{#2}}
\def\un#1,#2,#3~{${\rm u}_#1(#2,#3)$}
\def\mun#1,#2,#3~{{\rm u}_#1(#2,#3)}
\def\gQ#1,#2~{$g_\rQ(#1,#2)$}
\def\mgQ#1,#2~{g_\rQ(#1,#2)}
\def\taQ{$\tau^\rQ$}
\def\mtaQ{\tau^\rQ}
\def\mtQ#1,#2~{\tau^\rQ_{#1}#2}
\def\tQ#1,#2~{$\tau^\rQ_{#1}#2$}

\def\mv#1{{\bf v}_{#1}}
\def\w#1{${\bf w}_{#1}$}
\def\mw#1{{\bf w}_{#1}}
\def\vv#1,#2~{${\mathbf v}_{#1}(#2)$}
\def\mvv#1,#2~{{\mathbf v}_{#1}(#2)}
\def\vfn{${\mathbf v}_f(\nu)$}
\def\vvfn{${\mathbf{\check{v}}}_f(\nu)$}
\def\mvvfn{{\mathbf{\check{v}}}_f(\nu)}
\def\mvfn{{\mathbf v}_f(\nu)}
\def\vfr{${\mathbf v}_f(\mrh)$}
\def\Tpq{$\mcl T^p_q(M)$}
\def\Tpqx{$T^p_{qx}(M)$}
\def\mTpqx{T^p_{qx}(M)}
\def\mTpq{\mcl T^p_q(M)}
\def\mrTpq{T^p_q(M)}
\def\rTpq{$T^p_q(M)$}
\def\XN{$\mcl X(N)$}
\def\mXN{\mcl X(N)}
\def\XM{$\mcl X(M)$}
\def\mXM{\mcl X(M)}
\def\L#1~{$\pounds_{#1}$}
\def\mL#1~{\pounds_{#1}}
\def\dom{$\rd\mome$}
\def\mdom{\rd\mome}
\def\ip#1{$\boldsymbol{i_{#1}}$}
\def\mip#1{\boldsymbol{i_{#1}}}

\def\Lq{$L^2({\Bbb R},dq)$}
\def\mLq{L^2({\Bbb R},dq)}
\def\Wa{$W^*$-algebra}
\def\Wsa{$W^*$-subalgebra}
\def\Ca{$C^*$-algebra}
\def\Csa{$C^*$-subalgebra}
\def\rep{${}^*$--representation}
\def\Wrep{$W^*$--representation}
\def\autm{${}^*$-automorphism}
\def\aut#1{${}^*$-Aut\ #1}
\def\maut#1{{}^*$-Aut$\ #1}
\def\Aut#1{$Aut(#1)$}
\def\mAut#1{Aut(#1)}

\def\Psib{$\Psi^\flat$}
\def\mPsib{\Psi^\flat}

\def\lb{\langle}
\def\rb{\rangle}

\def\plr{$p_{\leftrightarrow}$}
\def\mplr{p_{\leftrightarrow}}

\def\eequiv{\Leftrightarrow}
\def\imply{\Rightarrow}

\def\wrt{with respect to\ }
\def\om#1,#2~{$\omega_{#1}^{#2}$}
\def\mom#1,#2~{\omega_{#1}^{#2}}
\def\ommr{$\mome_{\mu,\hat\mrh}$}
\def\mommr{\mome_{\mu,\hat\mrh}}

\def\prob{{\rm prob}}
\def\nbhd{neighbourhood\ }

\def\eq{{equation}}

\def\noidt{\noindent}

\def\emm#1~{{\bf #1}\ind{#1}}
\def\glss#1~{#1\glo{#1}}

\newcommand{\beq}{\begin{equation}}
\newcommand{\eeq}{\end{equation}}

\newcommand{\barr}{\begin{eqnarray}}
\newcommand{\earr}{\end{eqnarray}}

\newcommand{\glo}{\glossary}
\newcommand{\ind}{\index}
\newcommand{\rarw}{\rightarrow}
\renewcommand{\Im}{{\rm Im}}
\renewcommand{\Re}{{\rm Re}}

\swapnumbers \theoremstyle{plain}
\newtheorem{thm}{Theorem}[section] %%%\zal
\newtheorem{defi}[thm]{Definition}    %%%%\pika
\newtheorem{defs}[thm]{Definitions}   %%%%%\pika
\newtheorem{notat}[thm]{Notation}     %%\pika
\newtheorem{prop}[thm]{Proposition}    %%%%\zal
\newtheorem{lem}[thm]{Lemma}        %%%%%\zal
\newtheorem{lem*}[thm]{Lemma*}       %%%%\zal
\newtheorem{pt}[thm]{}             %%%%%\zel
\theoremstyle{remark}
\newtheorem{intpn}[thm]{\bf Interpretation}  %%%\bpika
\newtheorem{rem}[thm]{{\bf Remark}}         %%%%\dovi
\newtheorem{exmp}[thm]{{\bf Examples}}      %%%%%\dovi
\newtheorem{exm}[thm]{{\bf Example}}         %%\dovi
\newtheorem{ill}[thm]{{\bf Illustration}}    %%%%%%\dovi
\newtheorem{note}[thm]{{\bf Notes}}       %%%%%\dovi
\newtheorem{noti}[thm]{{\bf Note}}         %%%\dovi
\newtheorem{conj}[thm]{{\bf Conjecture}}  %%%\bpika
\newtheorem{corol}[thm]{{\bf Corollary}}  %%%\bpika

\def\EQM{\cite{bon}}
%%%%%%%%%%%%%%%%%%%%%%%%
\title{Some considerations on topologies of infinite dimensional
unitary coadjoint orbits.\footnote{This is a revised version of the preceding versions of this
paper. The main difference consists of the correction of a wrong assertion of the Lemma 3.1 based
on an erroneous assumption. The error was discovered in \cite[Sec. 5]{GKMS}.}}
%%%%%%%%%%%%%%%%%%%%%%%%%%%%
 \author{Pavel B\'ona \\ e-mail:\ bona@sophia.dtp.fmph.uniba.sk \\
Department of Theoretical Physics, Comenius University \\ SK-842 48 Bratislava, Slovakia}
%%%%%%%%%%%%%%%%%%%%%%%%%%%%%%%%%
\maketitle
\begin{abstract}{The topology of the embedding of the coadjoint orbits
of the unitary group \UH\ of an infinite dimensional complex Hilbert space \H, as canonically
determined subsets of the B-space \fTs\ of symmetric trace class operators, is investigated. The
space \fTs\ is identified with the B-space predual of the Lie-algebra \LHs\ of the Lie group \UH.
It is proved, that orbits consisting of symmetric operators with finite rank are (regularly
embedded) closed submanifolds of \fTs. An alternative method of proving this fact is given for the
``one-dimensional'' orbit, i.e. for the projective Hilbert space \PH. Also a technical assertion
concerning existence of simply related decompositions into one-dimensional projections of two
unitary equivalent (orthogonal) projections in `generic relative position' is formulated, proved,
and illustrated.}
\end{abstract}

%%%%%%%%%%%%%%%%%%%%%%%%%%%%%%%%%%%%%%%%%%%%%%%%%%%%%%%%%%
%\pagebreak
%%%%%%%%%%%%%%%%%%%%%%%%%%%%%%%%%%%%%%%%%%%%%%%%%
%\tableofcontents
%%%%%%%%%%%%%%%%%%%%%%%%%%%%%%%%%%%%%%%%%%%%%%%%%%

\hspace{1cm}
%\vspace*{\fill}

%\newpage
\section{Introduction}\label{sec;introd}
Mathematical formalism of non--Einstein-relativistic quantum mechanics (QM) is traditionally based
on separable complex Hilbert space \H, and on closely connected objects: The \Ca\ of bounded
operators \LH, the $\sigma(\mLH,\mLH_*)$--continuous (with $\mLH_*:=\mfk T:=L^1(\mH):=$ the
trace--class operators on \H) linear functionals on \LH\ (identified with $\nu\in\mfk T$), and the
group of $*-$automorphisms \aut(\LH)\ of \LH\ (acting on linear functionals by the transposed
maps). Dynamics (i.e. time evolution) and symmetries of physical systems are described by
subgroups of the automorphism group of \LH. Since each automorphism $\alpha\in\maut(\mLH)$ of the
\Ca\ \LH\ is inner, it is described by a unitary operator (\UH\ is the set of all unitary elements
of \LH)\ $\ru_\alpha\in\mUH: \alpha(B)=\ru_\alpha B \ru_\alpha^*$\ ($\ru_\alpha$\ is determined by
\alp\ up to a numerical factor). In physics, symmetries and dynamics are modelled by Lie groups
$G$. In the traditional {\em linear} QM, Lie groups are represented by strongly continuous unitary
(or projective) representations $g\ (\in G)\mapsto U(g)\ (\in\mUH)$, hence linear dynamics is
described by one parameter unitary groups $U_t\equiv \exp(-itH)$, with a selfadjoint operator $H$
on (a dense domain of) \H. Since physically interesting objects describing ``states'' are not
vectors $x\in\mH$, but to the vectors $x$ corresponding one-dimensional projectors $P_x$ onto the
subspaces of \H\ containing $x$, as well as their convex combinations (so called {\em mixed
states} described by `density matrices'
$\mrh:=\sum_j\lambda_jP_{x_j}\in\mSs:=\{\rho\in\mfTs:\rho\geq 0, Tr\rho=1\} \subset\mfTs$) the
physically interesting orbits of actions of the considered groups (resp. their representations)
are orbits of the coadjoint action of \UH\ (and of its subgroups), considered as a Lie group (see
below).

In a more general (also {\em nonlinear}) setting, cf. e.g. \EQM, symmetries and dynamics in such
an ``extended quantum mechanics'' (EQM) can be described by unitary cocycles $(g;\rho)\ (\in
G\times\mSs)\mapsto\mun Q,g,\rho~\ (\in\mUH),\ \mun Q,g_1\cdot g_2 ,\rho~=\mun
Q,g_1,\phi_{g_2}^{\rm Q}(\rho)~\cdot\mun Q,g_2,\rho~$, acting on the set \Ss\ of all `density
matrices', again by means of the coadjoint action of \UH, i.e. $\phi^{\rm Q}:\ (g;\rho)\ (\in
G\times\mSs)\ \mapsto \phi_g^{\rm Q}(\rho):=\mun Q,g,\rho~\cdot\rho\cdot\mun Q,g,\rho~^{-1}$ (here
Q is, in the case of one-parameter group $G:=\mbR$, a Hamiltonian function given on a Poisson
manifold, specifying the cocycle ${\rm u}_{\rQ}$). These actions leave the orbits of the coadjoint
representation $Ad^*(\mUH)$ going through $\rho\in\mSs$ again invariant, leaving invariant also
the whole set \Ss. This EQM is a general scheme of theories including Hamilton classical mechanics
(CM), linear QM, and also various versions of nonlinear QM, and also other in physics used
theoretical schemes as, e.g., various nonlinear ``approximations'' to QM (e.g., the time dependent
Hartree-Fock theory).

These remarks have to stress that the coadjoint orbits of the Lie group (see below) \UH\ going
through symmetric trace class operators are important mathematical objects in physical description
of a rather large scale of ``processes''.

As I have learned from a discussion with colleagues Anatol Odzijewicz and Tudor Ratiu, there is an
``innocently looking'' question connected with a work with coadjoint action of Lie groups, which
is far not trivial in the general case. It is the question in which way the  homogeneous spaces
$G/G_\rho$ of a Lie group $G$ with their natural analytic manifold structure (with $G_\rho$ being
the stability subgroup of $G$ at the point \rh), specifically their coadjoint orbits, are included
into the topological spaces where the group acts. In more specific terms the question is, whether
the injective inclusion is an immersion and homeomorphism of the analytic manifold $G/G_\rho$ onto
a submanifold of the space \cT\ on which the group $G$ acts. E.g., an orbit \cl O\ of a specific
action of \bR\ on the two-torus $T^2 = S^1\times S^1$, i.e. $\mcl O:=
\{(e^{it\omega_1};e^{it\omega_2}):t\in\mbR\}\subset T^2$ with irrational quotient
$\omega_1/\omega_2$, covers the torus densely, hence it is not a submanifold of $T^2$. As it is
shown in a Kirillov's example \cite{kiril} (cited and reproduced in \cite[14.1.(f),
p.449]{mars&rati}), such a pathologically looking case is possible also in the cases of
finite-dimensional {\em coadjoint orbits}. The more one could expect such a phenomenon in the case
of infinite-dimensional orbits of Banach Lie groups.

Let $\mOUr=\mfk U/\mUr$ be the homogeneous space of the unitary group \fk U := \UH\ of the
infinite-dimensional Hilbert space \H\ corresponding to an orbit of the action $\ru\mapsto \ru\rho
\ru^*$ on the space $\mfTs(\ni\rho)$\ of symmetric trace class operators in \LH. The space
$\mfTs$\ is naturally identified with the predual ${\mLH_s}_*$ of the Lie algebra $Lie\
\mUH:=i\mLH_s\sim\mLH_s$ (\Ur\ is the stability subgroup of \fk U\ at \rh, namely
$\mUr=\{\mrh\}'\cap\mfk U$, $\mrh^*=\mrh\in\mfTs$, with $\{A\}'$ being the commutant in \LH\ of
$A$).

In the paper \EQM, the topology of the orbits \OUr, as well as the topology of their natural
injection into the dual B-space (containing the predual \fTs) were investigated.\footnote{Let us
note that a far reaching generalization of some of structures developed and investigated in \EQM\
is contained in very elegant paper \cite{Odz&Rat}; that paper was also stimulating for the here
reported research.} It was proved there (cf. \cite[Proposition 2.1.5]{bon}), that orbits trough
symmetric trace-class operators are injectively immersed into \fTs\ iff they are going trough
operators with finite rank. There was not completed, however, the proof of an assertion on {\em
regularity} of this embedding (in the terminology of \cite{3baby}) of such ``finite-rank'' orbits,
which claim was contained in the text of the Proposition 2.1.5.\footnote{The claim of ``regularity
of the embedding'' was, however, superfluous (nevertheless correct, as it could be seen from what
we shall prove here) with respect to the validity of the Proposition 2.1.5 (without the
requirement of {\em regularity} of the embedding in its item (iv)) , as well as \wrt\ its actual
applications in all the paper \EQM.} One of the aims of this paper is to fill this gap.

Let us note, that the posed question of whether the orbit is also a submanifold of the ``ambient''
space in which the group acts is easily and positively answered in the case of finite-dimensional
Hilbert space \H: In that case the group \UH\ is compact, so that the orbits are also compact and
a continuous bijection of any compact space into a Hausdorff space is a closed mapping, hence a
homeomorphism. For an infinite-dimensional \H, however, the orbits \OUr\ are noncompact.

The proof of the main theorem is contained in the next Section \ref{sec;proof}. The presented
proof is based on a simple idea, and it does not contain any ``sophisticated mathematics''; it
needed just some linear algebra and elementary topology to be presented in details. In the last
Section \ref{sec;altern} some additional facts (including a proof of the fact that the orbits
consisting of finite-rank density matrices are closed subsets of the ``ambient'' space) are
presented. Also an independent proof of {\em regularity} of the embedding (we use here the
definitions adopted from \cite{3baby} differing from those introduced in \cite{mars&rati}, also to
keep continuity with \EQM) of the projective Hilbert space is presented: It indicates also an
alternative way for proving the main Theorem \ref{thm;reg-emb} for the general case.
%%%%%%%%%%%%%%%%%%%%%%%%%%%%%%%%%%%%%

\section{A proof of regularity of the embedding}\label{sec;proof}
 We shall accept here some conventions and results from \EQM, mainly
from the proof of Proposition 2.1.5 and Theorem 2.1.19. The proof of the following Theorem
\ref{thm;reg-emb}\ completes the missing part of the proof of Proposition 2.1.5 in \EQM\ concerned
the {\em regularity} of the embedding of $\mOUr\subset\mfTs$. Some of the constructions built and
used in the run of the proof might be, perhaps, also of independent interest.

Let us describe first in more details a formulation of the problem, and our strategy to approach
it. It is known, \cite[Proposition 37, Chap. III.\S 3]{bourb;Lie}, that the unitary group \fk U :=
\UH\ of the $C^*-$algebra \LH\ of all bounded operators on a complex Hilbert space \H\ is a Banach
Lie group, and its Lie algebra $Lie(\mfk U)$\ consists of all antisymmetric bounded linear
operators $i\mLHs$, which is B-space isomorphic to \LHs. The adjoint representation of \fk U\ in
the B-space $\mLHs$ is $Ad:\mfk U\rarw\mcl L(\mLHs),\ \ru\mapsto Ad(\ru)$, with $Ad(\ru)B:=\ru
B\ru^*,\ \forall B\in\mLHs$. The representation we are mostly interested in here is the {\em
coadjoint representation} consisting of the transposed mappings $Ad^*(\ru):=Ad(\ru^{-1})^*$ to
$Ad(\ru^{-1})$'s, hence acting on continuous linear functionals $\nu\in\mLHs^*,\ \nu:\mLHs\rarw
\mathbb C,\ B\mapsto \lb \nu;B\rb $; the mapping $Ad^*(\ru): \mLHs^*\rarw \mLHs^*$ is determined
by $\lb Ad^*(\ru)\nu;B\rb:=\lb\nu;Ad(\ru^{-1})B\rb$. The subset of symmetric {\em normal} linear
functionals can be identified with the B-space $\mfTs\subset\mLHs^*$ of symmetric trace-class
operators : $\nu\ (\in\mfTs):B\mapsto \lb\nu;B\rb:=Tr(\nu B)$; the space of normal (i.e.
continuous in the topology on \LH\ given by the seminorms $p_\nu: B\mapsto p_\nu(B):=|Tr(\nu B)|,\
\nu\in\mfTs$) symmetric functionals is a Banach space \fTs\ with the trace-norm
$\|\nu\|_1:=Tr|\nu|$, with the absolute value of the operator $\nu$\ defined as the operator
$|\nu|:=\sqrt{\nu^*\nu}\in\mLH$.

We are interested in comparison of two topologies introduced on the orbits $\mOUr:=\{\ru\rho
\ru^*:\ru^{-1}=\ru^*\in\mfk U\subset\mLH\}$ of the coadjoint representation. Let us denote
$\mUr:\{\ru\in\mfk U:\ru\rho=\rho \ru\}\ (\rho\in\mfTs)$. Then \Ur\ is a Lie subgroup of \fk U,
\cite[Lemma 2.1.2]{bon}, and the factor-space $\mfk U/\mUr$ (which can be canonically identified,
as a set, with \OUr) endowed with the factor-topology of the analytic Banach Lie group \fk U\ is
an analytic Banach manifold, \cite[III.1.6, Proposition 11]{bourb;Lie}. On the other side, the
orbit \OUr\ is naturally a subset of the Banach space \fTs\ endowed with the norm-topology given
by the trace-norm $\|\cdot\|_1$. The topology induced on \OUr\ from this B-space topology on \fTs\
need not coincide with the analytic manifold topology of \fk U/\Ur. It is known that this
coincidence is {\em not} the case for any \rh\ with infinite-dimensional range, cf.
\cite[Proposition 2.1.5]{bon}. The coincidence of these two topologies means that the immersed
subset $\iota(\mfk U/\mUr)=\mOUr$\ of \fTs\ endowed with the topology of \fk U/\Ur\ is a
submanifold of \fTs, or equivalently, that the inclusion mapping $\iota:\mfk U/\mUr\rarw\mfTs$
(provided that $\iota$ is immersion) is a homeomorphism of \fk U/\Ur\ onto the topological
subspace $\mOUr\subset\mfTs$, \cite[5.8.3]{bourb;manif}.

 We intend to prove that, for any $\rho=\rho^*\in\mfk F$\
(:= the linear space of finite-rank operators in a complex Hilbert space \H), the topology induced
on the subset $\mOUr:=\{\ru\rho \ru^*:\ru^{-1}=\ru^*\in\mfk U\subset\mLH\}$ from the overlying
(resp. ``ambient'') Banach space of symmetric trace-class operators \fTs\ is equivalent to the
topology of the set \OUr\ considered as the factor-space \fk U/\Ur. If the inclusion $\iota:\mfk
U/\mUr\rarw\mOUr\subset\mfTs,\ [\ru]_\mrh\mapsto \iota([\ru]_\mrh):=\ru\rho \ru^*$, where
$[\ru]_\rho:=\{\rv\in\mfk U:\rv\rho \rv^*=\ru\rho \ru^*\}$ is an (injective) immersion, and if it
were also homeomorphism of \fk U/\Ur\ onto $\iota(\mfk U/\mUr)=\mOUr$, then \OUr\ would be a
submanifold of \fTs, cf. \cite[5.8.3]{bourb;manif}.

Let us sketch our ``strategy'' of proving this claim here. It was proved in \cite[Proposition
2.1.5]{bon}, that \OUr\ is an immersed submanifold (i.e. the inclusion $\iota:\mfk
U/\mUr\rarw\mOUr\subset\mfTs$ is an immersion, \cite[5.7.1]{bourb;manif}) of \fTs\ for
$\dim(\rho):={\rm rank}(\rho)<\infty$. We are going to prove that the inverse mapping
$\iota^{-1}:\mOUr\rarw\mfk U/\mUr$ is also continuous. It will be useful to our technique to use
the metric-space description of continuity of mappings, i.e. the ``$\epsilon \leftrightarrow
\delta$ language''. It is useful to realize for this that the considered orbits \OUr\ are all (for
$\dim(\rho)<\infty$) Riemann manifolds endowed with strong riemannian metrics, \cite[Theorem
2.1.19]{bon}. Then the manifold topology is given by the corresponding distance function,
\cite[Proposition 4.64]{jt-schw}, hence all the considered topologies are metric ones, i.e. the
topology on \fk U\ given by the operator norm $\|\ru-\rv\|$, the riemannian topology on \OUr\
represented by a distance function $d_\rho(\rho',\ru\rho'\ru^*)$,\footnote{The distance $d_\rho$
will not be explicitly calculated here.} and also the topology of the space \fTs\ into which is
\OUr\ embedded is given by the norm $\|\rho'-\ru\rho'\ru^*\|_1$.

We have to prove that, for any $\rho'\in\mOUr$, and for an arbitrary (small) $\epsilon'>0$ there
is a $\delta'>0$ such that if there is an element $\rho''\in\mOUr$ with
$\|\rho''-\rho'\|_1<\delta'$, then it is also $d_\rho(\rho',\rho'')<\epsilon'$. The projection
$\Pi_\rho:\mfk U\rarw\mOUr,\ \ru\mapsto[\ru]_\rho\approx \ru\rho \ru^*$ is continuous (here
$[\ru]_\rho\approx \ru\rho \ru^*$ means the canonical identification of the left cosets
$[\ru]_\rho\subset\mfk U$ with their realization as the points $\ru\rho \ru^*$ of \OUr). We can
use this continuity to avoid necessity of (possibly complicated) calculation od explicit forms of
$d_\rho$, cf., however, Proposition \ref{prop;PH}: Since $\Pi_\rho$ is uniformly continuous (due
to obvious invariance of both metrics), to any $\epsilon'>0$ there is an $\epsilon>0$ such that if
$\|\ru-\rv\|<\epsilon$, then also $d_\rho(\ru\rho \ru^*,\rv\rho \rv^*)<\epsilon'$. So, if we could
find to any $\epsilon>0$ and a $\rho'\in\mOUr$ such a $\delta'>0$ that for any $\rho'':=\ru\rho'
\ru^*: \|\rho'-\rho''\|_1<\delta'$ \emph{it is possible to find} an unitary $\rv$ such that also
$\rho''=\rv\rho'\rv^*$, and such that also $\|I_\mH-\rv\|<\epsilon$, then the continuity will be
proved. We shall proceed essentially in this way, but to avoid explicit calculation of dependence
$\epsilon\mapsto \delta'(\epsilon)$, we shall use also another known continuity, namely the
continuous dependence of the spectral projections $F_j(\rho'')$ of
$\rho'':=\sum_j\lambda_jF_j\in\mOUr$ onto the $\rho''$ itself. Also the homogeneity of the orbit
and of its ``ambient'' space \fTs\ will lead to a simplification.

The following lemma provides a reader with a `freedom' in dealing with various topologies induced
on the considered orbits.
\begin{lem}
The topologies coming from the trace class B-space $L^1(\mH)(:=\mfk{T}(\mH)\supset
{\mfTs}\supset{\mfk{F}}_N)$, from the Hilbert-Schmidt B-space
$L^2(\mH)(:=\mfk{H}\supset\mfk{H}_s\supset {\mfTs}\supset{\mfk{F}}_N)$, as well as from the \Ca\
of all bounded operators $L^\infty(\mH):=\mLH(\supset\mLHs\supset\mfk{H}_s\supset
{\mfTs}\supset{\mfk{F}}_N)$, induced on the subset of symmetric finite rank operators
${\mfk{F}}_N$ with a fixed maximal dimension $N$ of their ranges are all equivalent.\zal
\end{lem}
\begin{proof} These topologies are equivalent in finite dimensional
linear spaces. Explicitly, in our case: Let $N$ be maximal dimension of ranges of the considered
operators $A,B\in{\mfk{F}}_s,\ A=A^*,\ B=B^*,$ hence the ranges of the operators $A-B$ are of
maximal dimension $2N$. The considered topologies are all metric topologies induced on
${\mfk{F}}_N$ by the corresponding norms from the ``above lying'' spaces. The distances between
$A$ and $B$ are correspondingly given by $\|A-B\|_1:=Tr|A-B|,\ \|A-B\|_2:=\sqrt{Tr|A-B|^2}$, and
$\|A-B\|=:\|A-B\|_\infty$\ =\ {\em the maximal eigenvalue of}\ $|A-B|$, where $|A-B|$ denotes the
absolute value of the operator $A-B$, $|A-B|:=\sqrt{(A-B)^*(A-B)}$. Generally it is
$\|C\|_\infty\equiv\|C\|\leq\|C\|_2\leq\|C\|_1$ for any trace-class operator $C$. Conversely, also
due to the mentioned inequalities, one clearly has $\|A-B\|_2\leq\|A-B\|_1\leq
2N\|A-B\|_\infty\leq 2N\|A-B\|_2$ for $A, B\in{\mfk{F}}_N$. This shows that all the three metric
topologies are on ${\mfk F}_N$ mutually equivalent.\end{proof}

We shall need a rather indirect, but a quite ``faithful'' expression for ``proximity'' of
finite-rank operators on the same orbit considered as a subset of the B-space \fTs, which would be
more difficult to express directly with a help the usual norms of their differences. To this end
we shall need the following lemma.
\begin{lem}\label{lem;contproj}
Let us consider a subset ${\mfk F}_\sigma$ of bounded symmetric operators $\mrh\in\mLH$\ with a
given purely discrete finite spectrum $\sigma:=\{\lambda_0,\lambda_1,\lambda_2,\dots
\lambda_n\}\subset\mbC$. Their spectral projections $F_j\equiv F_j(\rho), (j=0,1,2,\dots n)$ are
continuous functions of $\mrh\in\mfk F_\sigma$:
$$ \mrh:=\sum_{j=0}^n \lambda_j\cdot F_j,$$
in the operator norm topology of \LH.\zal
\end{lem}
\begin{proof} The spectral projections of any symmetric operator \rh\ are
uniquely determined by that operator, hence for a given spectrum (e.g. $\mrh\in\mfk F_\sigma$) the
projections corresponding to fixed spectral values are uniquely determined functions of the
operators $\mrh\in\mfk F_\sigma$. By a use of a spectral functional calculus one can choose some
functions $p_j:\mbR\rarw\mbR$ such, that $p_j(\lambda_k)\equiv \delta_{jk}$. Then
$p_j(\rho)=F_j:=F_j(\rho), \forall j$. Let us choose for the functions $p_j$ polynomials; we
define for any complex $z\in\mbC$

\beq p_j(z):=\prod_{k(\neq j)=0}^n \frac{z-\lambda_k}{\lambda_j-\lambda_k},\eeq what gives
$p_j(\rho)=F_j(\rho)$, and the continuity of $\mrh\mapsto F_j(\rho)$ on (any subset of) $\mfk
F_\sigma$ is explicitly seen.\end{proof} \noidt This two Lemmas lead immediately to

\begin{corol}
The spectral projections $F_j$ of finite rank operators $\rho\in\mfk F_\sigma\cap\mfk F_N$ are (on
the set $\mfk F_\sigma\cap\mfk F_N$) continuous functions $\rho\mapsto F_j(\rho)$ of these
operators in any of the considered (i.e. trace, Hilbert-Schmidt, and \LH) topologies (taken
independently on the domain-, or range-sides). \bpika\end{corol}

 We shall use in the following text also the Dirac notation for
vectors and operators in a complex Hilbert space: $|x\rb:=x\in\mH$ will denote a vector, $\lb
x|y\rb$ is the scalar product of such vectors (linear in the {\em second} factor), and $|x\rb\lb
y|$ is the operator of one-dimensional range such, that $|x\rb\lb y|:\sum_jc_j|z_j\rb\mapsto
|x\rb\lb y|\cdot\sum_jc_j |z_j\rb:=\left(\sum_jc_j\lb y|z_j\rb\right) |x\rb$.

 The constructions needed in the proof of the main theorem use
also a more detailed description of consequences of ``proximity'' of two projections described in
the following

\begin{lem}\label{lem;proxi}
Let $E,\ F$\ be two orthogonal projections of finite-dimensional ranges of equal dimensions
$N:=\dim E=\dim F:=Tr(E)$ in an infinite-dimensional Hilbert space \H. Assume that $E\wedge F=0$,
i.e. the subspaces $\mcl E:=E\mH$ and $\mcl F:=F\mH$ have no nonzero common vectors. Let us also
denote $\mcl E\vee\mcl F:=\mcl E+\mcl F=(E\vee F)\mH$ the $2N$-dimensional linear hull in \H\ of
$\mcl E\cup\mcl F$. Let

\beq\label{eq;EF} Tr[(E-F)^2]\equiv\|E-F\|^2_2<2. \eeq

Then:\item{(i)} For any one-dimensional projections given by normalized vectors $e\in\mcl E,\
f\in\mcl F: |e\rb\lb e|=:P_e\leq E\ (i.e.\ P_e\cdot E=P_e)$, and $|f\rb\lb f|=:P_f\leq F$,  it
is:\quad $P_e\cdot F\neq 0$, and $P_f\cdot E\neq 0$.

\item{(ii)} There exists an orthonormal basis
$\{e_j:j=1,2,\dots, N:=\dim E\}\subset \mH$ in \cl E, i.e. $\sum_j P_{e_j}=E$, such that one can
find to it an orthonormal basis of \cl F:\ $\{f_j:j=1,2,\dots, N\}\subset\mH$ (i.e. $\sum_j
P_{f_j}=F$), satisfying the relations

\beq\label{eq;proj}  P_{f_j}(E-P_{e_j})=0,\quad P_{e_j}(F-P_{f_j})=0,\quad \forall j. \eeq

\item{(iii)} This means
that these orthonormal systems $\{e_j:j=1,2,\dots, N\}$, and $\{f_j:j=1,2,\dots, N\}$, decomposing
$E$ and $F$, are in a certain strong sense mutually ``affiliated'':

\beq\label{eq;Fproj} F|e_j\rb= |f_j\rb\lb f_j|e_j\rb,\quad \forall j=1,2,\dots N,\ 0\neq\lb
f_j|e_j\rb\in\mbC,\ \lb e_j|e_j\rb\equiv 1\equiv\lb f_j|f_j\rb,
 \eeq
i.e. from a specific orthonormal `decomposition' $\{e_j:j=1,2,\dots, N\}$ of $\mcl E$ the
orthonormal system $\{ f_j:j=1,2,\dots, N\}$ `decomposing' $\mcl F$ and satisfying \rref proj~ is
obtained, uniquely up to a nonzero numerical factor, simply by element-wise orthogonal projections
of $e_j$'s onto $\mcl F:=F\mH$.

\item{(iv)} The above mentioned specific orthonormal basis $\{e_j:j=1,2,
\dots, N\}$ determines also (up to `phase factors') an orthonormal basis $\{e_j^\bot:j=1,2,\dots,
N\}$ of $\mcl E^\bot:=[(E\vee F)-E]\mH=\mcl E\vee\mcl F\ominus\mcl E$, such that
$f_j=\alpha_je_j+\beta_je_j^\bot,\ \alpha_j\cdot\beta_j\neq 0,\ (\forall j).$
 \zal\end{lem}
\begin{proof}
\item{(i):} Let there be a projection $P_e\leq E$ such that $P_e F=0$.
Let $e_1:=e$, and let $\{e_j:j=1,2,\dots N\}$ be an orthonormal system decomposing $E,\
E=\sum_{j=1}^NP_{e_j}$. Then

\beq\label{eq;ii}Tr(EF)=Tr[(E-P_e)F]=\sum_{j=2}^NTr(P_{e_j}F)\leq N-1,\eeq

\noidt since always it is $Tr(P_xF)\leq 1, \forall x\in\mH$. The estimate \rref ii~ would be then
in contradiction with the assumption \rref EF~, since $Tr[(E-F)^2]=2(N-Tr(EF))$. Due to the
symmetry of the assumed conditions \wrt\ the exchange $E\leftrightarrow F$, one obtains also
$P_fE\neq 0$. This implies validity of (i).
\item{(ii):}
We have to prove existence of the bases $\{e_j\}:= \{e_j:j=1,2,\dots N:=\dim E\}$, and
$\{f_j:j=1,2,\dots N=\dim F\}$ of \cl E, resp. \cl F\ satisfying \rref proj~.

This means to find an orthonormal basis $\{e_j:j=1,2,\dots N\}$ of \cl E\ such, that its
element-wise projections are proportional to $f_j$'s, cf. \rref Fproj~. This also means that for
such a basis $\{e_j\}\subset\mcl E$ the projections $F|e_j\rb\in\mcl F$ are nonzero and mutually
orthogonal.

The statement (i) ensures that all the projections $F|e\rb$ of all nonzero vectors $e\in\mcl E$
are nonzero, i.e. that the restriction $EFE\in\mcl L(\mcl E)$ of the projector $F$ to the subspace
$\mcl E\subset\mH$  has trivial kernel: ${\rm Ker}_{\mcl E}(EFE)=0$. This implies that the bounded
operator $EFE=(FE)^*FE$ on \cl E\ is strictly positive and there is an orthonormal basis $\{e_j\}$
of \cl E\ in which the matrix $\lb e_j|EFE|e_k\rb=\lb e_j|F|e_k\rb$ is diagonal, with strictly
positive diagonal elements $\|Fe_j\|^2$.

Let us define then, e.g., $f_j:=\|Fe_j\|^{-1}\cdot F e_j,\ j=1,2,\dots,N$; these elements form the
wanted decomposition of \cl F, resp. of the projector $F$ satisfying together with the just found
basis $\{e_j\}$ the relations \rref proj~. This proves (ii).

\item{(iii):} That statement is just a rephrasing of (ii); the
uniqueness also is seen from \rref proj~.

\item{(iv):}  Since each $f_j\in\mcl F$ constructed as above is
orthogonal to all the $e_k (k\neq j)$, and $\lb f_j|e_j\rb\neq 0$, but it is also $E^\bot f_j\neq
0$, with $E^\bot:=E\vee F-E$, $f_j$ is expressible in the form

\beq\label{eq;e-bot} f_j:=\alpha_je_j+\beta_je_j^\bot,\ \forall j, \eeq where $e_j^\bot\in\mcl
E^\bot:=E^\bot\mH$ is some normalized vector determined by $f_j$ up to a `phase factor', e.g.:
$e_j^\bot:=\|E^\bot f_j\|^{-1}E^\bot f_j$.

We also see that all $\alpha_j\cdot\beta_j\neq 0$, since all $f_j\not\in \mcl E$, but also
$f_j\not\in\mcl E^\bot$.

The orthogonality between the vectors $f_j$'s : $\lb f_j|f_k\rb\equiv\delta_{jk}$ implies also the
orthogonality relations for $e_j^{\bot}$'s: $\lb e_j^\bot|e_k^\bot\rb=\delta_{jk}$.\end{proof}

The following lemma is an illustration of one of the main tools used in the proof of the
forthcoming theorem:

\begin{lem}\label{lem;unit}
Let $E, F$ be two orthogonal projections in an infinite-dimensional complex Hilbert space \H\ of
the same finite dimension $N=Tr(E)=Tr(F)$. Let us choose $0<\epsilon<2,$ and assume that
$N-Tr(EF)<\epsilon^2/4\ (<1).$ Then there is a unitary operator $\ru\in\mfk U$ such that
$\|\ru-I_\mH\|<\epsilon$, and that $F=\ru E\ru^*$.\zal
\end{lem}
\begin{proof} Let us denote $Q:=E\wedge F,\ E':=E-Q,\ F':=F-Q,\
N':=Tr(E')=Tr (F')=N-\dim Q,\ E'^\bot:=(E'\vee F')-E',\ F'^\bot:=(E'\vee F')-F'.$ Then $E'\wedge
F'=E'^\bot\wedge F'^\bot=0.$ We have now also $Tr(E'^\bot)=Tr(F'^\bot)=N'=:N'^\bot$. Moreover,

\barr\label{eq;delta}
 1>\epsilon^2/4>
N-Tr(EF)=\frac{1}{2}Tr[(E-F)^2]=\frac{1}{2}Tr[(E'-F')^2]=\nonumber\\
N'-Tr(E'F')= \frac{1}{2}Tr[(E'^\bot-F'^\bot)^2]=N'^\bot-Tr(E'^\bot F'^\bot). \earr

We can now apply Lemma \ref{lem;proxi} to the both couples, i.e. to $(E';F')$, as well as to
$(E'^\bot;F'^\bot)$, of projections. Let
 $\{e_j:j=1,2,\dots N'\}$, resp. $\{e_j^\bot:j=1,2,\dots N'\}$ be
 the orthonormal decompositions of $\mcl E':=E'\mH$, resp. of
 $\mcl E'^\bot:=[E'\vee
F'-E']\mH$, with the corresponding orthonormal decompositions $\{f_j:j=1,2,\dots N'\}$ of $\mcl
F':=F'\mH$, resp. $\{f_j^\bot:j=1,2,\dots N'\}$ of $\mcl F'^\bot:=F'^\bot\mH$ constructed
according to Lemma \ref{lem;proxi}. Because the normalized vectors $f_j, f_j^\bot$ remained
specified, according to Lemma \ref{lem;proxi}, up to arbitrary phase factors, we shall choose them
so that the scalar products $\lb f_j|e_j\rb>0,\ \lb f_j^\bot|e_j^\bot\rb>0$. Remember also that
for $j\neq k:\lb f_j|e_k\rb=0,\ \lb f_j^\bot|e_k^\bot\rb=0$. We have constructed two orthonormal
decompositions of the space $\mcl E'\vee\mcl F':= (E'\vee F')\mH$, i.e.
$\{e_j,e_j^\bot:j=1,2,\dots N'\}$, as well as $\{f_j,f_j^\bot:j=1,2,\dots N'\}$. We could also
define formally $\{e_j=f_j:j=N'+1,\dots N\}$ as an arbitrary orthonormal decomposition of
$Q=E-E'=F-F'$, but it will not be used now.

Let us define now the wanted unitary operator $\ru\in\mfk U$ by:

\beq \ru x:=x\quad {\rm for}\ x\in\mH\ominus(\mcl E'\vee\mcl F');\qquad \ru e_j:=f_j,\ \ru
e_j^\bot:=f_j^\bot\quad (j=1,2,\dots N'), \eeq

\noidt and this prescription is completed by linearity to a unique unitary operator $\ru$ on \H.
Let us prove that this operator has the wanted property. It is clear that $\ru E\ru^*=F$: $\ru
E\ru^*=\ru Q\ru^*+\ru E'\ru^*=Q+\sum_{j=1}^{N'}\ru|e_j\rb\lb e_j|\ru^*=
Q+\sum_{j=1}^{N'}|f_j\rb\lb f_j|= Q+F'=F.$ Since on the complement of the finite-dimensional
subspace $\mcl E'\vee\mcl F'$ of \H\ the operator $\ru$ coincides with $I_\mH$, their difference
$\ru-I_\mH$ can be nonzero just on the finite dimensional subspace $\mcl E'\vee\mcl F'$. Hence the
norm $\|\ru-I_\mH\|$ can be calculated as the norm of the restriction to the subspace $\mcl
E'\vee\mcl F'$, and we can deal with this operator $\ru-I_\mH$ as with a finite-dimensional
matrix. Or, the operator $\ru-I_\mH$ is of finite rank in \H. Let us denote $Tr'(C)$ the trace of
the restriction of $C\in\mLH$ to the $2N'-$dimensional subspace $\mcl E'\vee\mcl F' :\
Tr'(C):=Tr[(E'\vee F')C]$. We have

\barr\label{eq;unit}
 \|\ru-I_\mH\|^2\leq\|\ru-I_\mH\|^2_2=Tr'[(\ru^*-I_\mH)(\ru-I_\mH)]=
 Tr'[2I_\mH-\ru-\ru^*]=\nonumber\\
4N'-\sum_{j=1}^{N'}[\lb e_j|f_j\rb+\lb e_j^\bot|f_j^\bot\rb+\lb f_j|e_j\rb+\lb
f_j^\bot|e_j^\bot\rb]=\nonumber\\
4N'-2\sum_{j=1}^{N'}[|\lb e_j|f_j\rb|+|\lb e_j^\bot|f_j^\bot\rb|], \earr

\noidt due to the chosen positivity of the scalar products $\lb e_j|f_j\rb,\ \lb
e_j^\bot|f_j^\bot\rb$. According to \rref delta~, and also from the orthogonality properties of
the sets of chosen vectors $\{e_j,e_j^\bot,f_j,f_j^\bot:j=1,2,\dots N'\}$, and because it is $|\lb
e|f\rb|^2\leq |\lb e|f\rb|\leq 1$ for scalar product of any two normalized vectors $e,\ f$\ in \H,
one has

\barr 2N'-\sum_{j=1}^{N'}[|\lb e_j|f_j\rb|+|\lb e_j^\bot|f_j^\bot\rb|]\leq
2N'-\sum_{j=1}^{N'}[|\lb e_j|f_j\rb|^2+|\lb
e_j^\bot|f_j^\bot\rb|^2]=\nonumber\\
(N'-Tr(E'F'))+(N'^\bot-Tr(E'^\bot F'^\bot)<\frac{\epsilon^2}{2}. \earr

\noidt We have obtained, according to \rref unit~, the wanted estimate
$\|\ru-I_\mH\|^2<\epsilon^2$.\end{proof}

We are prepared now to prove the regularity of embeddings into \fTs\ of unitary orbits through
finite-rank symmetric operators.
\begin{thm}\label{thm;reg-emb}
Let $0\neq\rho=\rho^*\in{\mfk F}$\ (:=the set of all finite-rank operators on \H),
$\mOUr:=\{\ru\rho\ru^*:\ru\in\mfk U\}\subset\mfTs$. The unitary orbit \OUr\ is a regularly
embedded \cite[p. 550]{3baby} submanifold of the Banach space \fTs\ of symmetric trace--class
operators endowed with its trace norm.\zal
\end{thm}
\begin{proof}
%[``Projectors comparing'' form of the proof:]
The mapping $\Pi_\rho:\mfk U\rarw\mOUr, \ru\mapsto \ru\rho \ru^*$ is an analytic submersion
\cite[III.\S 1.6, Prop.11]{bourb;Lie}, and the inclusion $\iota_\rho:\mOUr\rarw\mfTs$ is an
injective immersion (cf. \cite[Proposition 2.1.5]{bon}), hence the composition
$\iota_\rho\circ\Pi_\rho,\ \mfk U\rarw\mfTs$ is continuous. We want to prove, that the inverse
(identity) mapping $\iota_\rho^{-1}:\mOUr\ (\subset\mfTs)\rarw\mOUr\ (:=\mfk U/\mUr)$ is also
continuous, if the ``domain copy'' \OUr\ of $\iota_\rho^{-1}$ is taken in the relative topology of
the corresponding ``ambient'' space $\mfTs\subset L^1(\mH)$. Because of the invariance of all the
relevant metrics \wrt\ the unitary group action (including their invariance in the ``ambient''
normed spaces), and also because of the continuity of the projection $\Pi_\rho$, it suffices to
prove the wanted continuity in an arbitrary point \rh\ of the orbit by showing the following: To
any positive $\epsilon>0$ one can find a $\delta'>0$ such, that if there is some element
$\mrh'=\ru\rho \ru^*\in\mOUr$ in the $\delta'-$\nbhd\ of \rh\ in the space ${\mfTs}:
\|\mrh-\ru\rho \ru^*\|_1<\delta'$, then {\em it is possible to find} a unitary $\rv\in\mfk
U:\|\rv-I_\mH\|<\epsilon$, such that $\rv\mrh\rv^*=\ru\mrh\ru^*$.

Now we can use, for the sake of simplicity of our expression, that the orbit \OUr\ is also a
strong riemannian manifold \cite[Thm. 2.1.19]{bon}  with a distance-function
$d_\rho(\rho',\rho'')$ generating the topology of\ \fk U/\Ur\ (cf.\ \cite[Proposition
4.64]{jt-schw}). Now (due to the continuity of $\Pi_\rho$), to any $\epsilon'>0$ there is an
$\epsilon>0$\ such that if $\|\rv-I_\mH\|<\epsilon$, then $d_\rho(\rho,\rv\rho \rv^*)<\epsilon'$.
We have to prove that, to this $\epsilon$, there exists the corresponding $\delta'>0$ such that
$\|\rho-\ru\rho \ru^*\|_1<\delta'\imply d_\rho(\rho,\ru\rho \ru^*)<\epsilon'$, what means the
desired continuity. {\em The proof will be direct: A construction of a unitary
$\rv:\|\rv-I\|<\epsilon$ for any given $\rho'=\ru\rho \ru^*$ lying ``sufficiently close'' to \rh\
in \fTs, such that it is also $\rho'=\rv\rho \rv^*$.}

Let us write $\rho=\sum_{j=1}^n \lambda_jE_j,\ 0<n<\infty$, where $\lambda_j\neq\lambda_k$ for
$j\neq k$, $E_j$ are the orthogonal projections of the spectral measure of $\rho=\rho^*$,\ $0<\dim
E_j:=Tr(E_j)=:N_j<\infty\ (\forall j\neq 0),\ E_0:=I_\mH-\sum_{j=1}^n E_j=:I_\mH-E,\
\lambda_0:=0,\ \sum_{j=1}^n N_j=:N$. Let us denote $F_j:=\ru E_j\ru^*\ (\forall j)$, hence
$\rho':=\ru\rho \ru^*=\sum_j\lambda_j F_j$, and also $F:=\sum_{j=1}^nF_j$.

It is clear that the nonnegative numbers $N_j-Tr(E_jF_j(\rho'))$ and $N-Tr(EF(\rho'))$ are all
continuous functions of $\mrh'$, and for $\rho'=\rho$ they are all zero. This can be seen, e.g. by
representing the projection operators $F_j\equiv F_j(\rho')$ by polynomials $p_j$ of the operators
$\rho'$, as it was done in Lemma\ \ref{lem;contproj}.

 These considerations imply that, for all sufficiently small
 $\delta'>0$, and for all such $\rho'=\ru\rho \ru^*$ that
$\|\mrh-\ru\mrh \ru^*\|_1<\delta'$, one obtains

\barr\label{eq;delta1} 0\leq N_j-Tr(E_jF_j(\rho'))=:\delta_j<1,\
j=1,2,\dots n;\nonumber\\
\ 0\leq N-Tr(EF(\rho'))=:\delta<1,
 \earr

\noidt where $\delta,\ \delta_j\ (j=1,2,\dots n)$ can be chosen arbitrarily small positive numbers
(i.e. they can be bounded from above by arbitrarily small positive upper bounds determining the
choice of the mentioned $\delta'>0$, what is possible due to the continuous dependence on $\rho'$
of the expressions entering into \rref delta1~).

Let us choose now $0<\epsilon<1$, and assume that the above mentioned $\delta'$ is
such\footnote{We need not here any explicit expression for the dependence
$\epsilon\mapsto\delta'\equiv\delta'(\epsilon)$; it could be `in principle' obtained, however,
from explicit formulas for the functions $\rho'\mapsto F_j(\rho')$, e.g. from those given in the
proof of Lemma \ref{lem;contproj}.} that

\beq\label{eq;delta2}\delta\leq\sum_{j=1}^n\delta_j<\epsilon^2/4,\eeq where the first inequality
is a consequence of the definitions \rref delta1~.

We shall now construct, for any $\rho'=\ru\rho \ru^*$ with $\|\rho'-\rho\|_1<\delta'$, such a
unitary $\rv\in\mfk U$, that $\rv\rho \rv^*=\ru\rho \ru^*$, and simultaneously
$\|\rv-I_\mH\|<\epsilon$.

 Let us
denote $Q_j:=E_j\wedge F_j,\ E_j':=E_j-Q_j,\ F_j':=F_j-Q_j,\ Q:=E\wedge F,\ E':=E-Q,\ F':=F-Q,\
E'^\bot:=(E'\vee F')-E'=E\vee F-E,\ F'^\bot:=(E'\vee F')-F'=E\vee F-F,\ N_j':=\dim E_j-\dim
Q_j=\dim E_j'=\dim F_j',\ N':=\dim E-\dim Q=\dim E'=\dim F'=\dim E'^\bot=\dim F'^\bot.$ Observe
that $(E-F)^2=[(E\vee F-E)-(E\vee F-F)]^2=(E'^\bot-F'^\bot)^2.$ Also it is
$Tr(EF)=Tr(E'F'+Q)=Tr(E'F')+N-N'$, and $\dim(E\vee F)=N+N'$. So that we obtain

\beq\label{eq;EFbot} Tr[(E-F)^2]=2[N-Tr(EF)]= Tr[(E'^\bot-F'^\bot)^2]=2[N'-Tr(E'^\bot F'^\bot)].
\eeq
   Now we can apply
Lemma\ \ref{lem;proxi} separately to each of the couples of projections

\beq\label{eq;pairs} (E_j';F_j'),\quad j=1,2,\dots n;\qquad\ (E'^\bot;F'^\bot),\eeq
 and construct the orthonormal systems
$\{e^{(j)}_k:k=1,2,\dots N_j'\}$ forming the convenient bases of every $\mcl E_j':=E_j'\mH\
(j=1,2,\dots n),$ and also the basis $\{e_k^\bot: k= 1,2,\dots N'\}$ of $\mcl
E'^\bot:=E'^\bot\mH$, such that their respective orthogonal projections onto the spaces $\mcl
F_j':=F_j'\mH\ (j=1,2,\dots n)$, and $\mcl F'^\bot:=F'^\bot\mH$, corresponding to the second
projection in the considered pair of \rref pairs~, are the orthogonal (and afterwards normalized)
bases
 $\{f^{(j)}_k:k=1,2,\dots
N_j'\}$ of $\mcl F_j'\ (j=1,2,\dots n)$, and the orthonormal basis $\{f_k^\bot: k= 1,2,\dots N'\}$
of $\mcl F'^\bot$. Let us choose any orthonormal bases $\{e^{(j)}_k\equiv f^{(j)}_k:
k=N'_j+1,\dots N_j\}$ of all the subspaces $\mcl Q_j:=Q_j\mH,\ j=1,2,\dots n$. We have obtained in
this way two orthonormal systems $\{e^{(j)}_k, e_i^\bot:\ k=1,2,\dots N_j,\ j= 1,2,\dots n,\
i=1,2,\dots N'\}$, and $\{f^{(j)}_k, f_i^\bot:\ k=1,2,\dots N_j,\ j= 1,2,\dots n,\ i=1,2,\dots
N'\}$, each forming a basis of the subspace $\mcl E\vee\mcl F:=(E\vee F)\mH$. Remember also the
``cross-orthogonality'' of the mutually ``affiliated'' orthonormal systems:

\beq\label{eq;orthog1} \lb f^{(j)}_k|e^{(j)}_l\rb = 0\quad(j=1,2,\dots,n),\qquad \lb
f^\bot_k|e^\bot_l\rb=0,\qquad {\rm for}\quad l\neq k \quad(\forall k,l). \eeq

Let also the arbitrary phase factors at the all $f$'s entering into the orthonormal sets be chosen
so that for all possible values of the indices it is

\beq\label{eq;posit} \lb f_l^\bot|e_l^\bot\rb>0,\qquad  \lb f^{(j)}_k|e^{(j)}_k\rb>0.\eeq

Now we shall define the wanted unitary $\rv$: Let the restriction of $\rv$ to $\mH\ominus(\mcl
E\vee\mcl F):=(\mcl E\vee\mcl F)^\bot$ be the identity (i.e. $\rv \rceil_{\mH\ominus(\mcl
E\vee\mcl F)}:= I_{\mH\ominus(\mcl E\vee \mcl F)}$), and its restriction to $\mcl E\vee\mcl F$ is
defined as the linear transformation between the constructed orthonormal systems forming two bases
in $\mcl E\vee\mcl F$ specified by:

\beq\label{eq;v} \rv e^{(j)}_k:=f^{(j)}_k,\qquad \rv e_i^\bot:=f_i^\bot;\qquad\forall i,j,k. \eeq

It is clear from this definition of $\rv$, esp. from \rref v~, that
$\sum_{j=1}^n\lambda_jF_j=\rv(\sum_{j=1}^n\lambda_jE_j)\rv^*$, i.e. $\rho'=\rv\rho \rv^*$. Let us
show next, that $\|\rv-I_\mH\|<\epsilon$. Since $(\rv-I_\mH)\rceil_{\mH\ominus(\mcl E\vee\mcl
F)}=0$, we shall estimate the Hilbert-Schmidt norm of $(\rv-I_\mH)$ in the subspace $\mcl
E\vee\mcl F$. Let $Tr'(C)$ will be the trace of the restriction of $C\in\mLH$ to $\mcl E\vee\mcl
F$. We obtain with a help of \rref posit~:

\barr \|\rv-I_\mH\|_2^2=  Tr'(2I_\mH-\rv-\rv^*)=\qquad\nonumber\\
  2(N+N')-2\sum_{j=1}^n\sum_{k=1}^{N_j}\lb
f^{(j)}_k|e^{(j)}_k\rb-2\sum_{j=1}^{N'}\lb
f_j^\bot|e_j^\bot\rb=\qquad\nonumber\\
  2\sum_{j=1}^n[N_j-\sum_{k=1}^{N_j}\lb
f^{(j)}_k|e^{(j)}_k\rb]+2[N'-\sum_{j=1}^{N'}\lb f_j^\bot|e_j^\bot\rb]\leq\qquad
\nonumber\\
  2\sum_{j=1}^n[N_j-\sum_{k=1}^{N_j}|\lb
f^{(j)}_k|e^{(j)}_k\rb|^2]+2[N'-\sum_{j=1}^{N'}|\lb
f_j^\bot|e_j^\bot\rb|^2]=\qquad\nonumber\\
  2\sum_{j=1}^n[N_j-Tr(E_jF_j)]+2[N'-Tr(E'^\bot
F'^\bot)]=\qquad\nonumber\\
  2\sum_{j=1}^n[N_j-Tr(E_jF_j)]+2[N-Tr(EF)],\qquad
 \earr

\noidt where we have used again the orthogonality properties \rref orthog1~\ of the vectors inside
each ``block'' corresponding to $E_j,\ j=1,2,\dots n$, as well as to $E'^\bot:\
\sum_{j=1}^nE_j+E'^\bot=E\vee F$, the fact that $|\lb f|e\rb|^2\leq|\lb f|e\rb|$ for any
normalized vectors $e,f\in\mH$, and also the relation \rref EFbot~.

Now we shall use the definitions \rref delta1~, and the assumption \rref delta2~. We obtain:

\beq \|\rv-I_\mH\|^2\leq\|\rv-I_\mH\|_2^2\leq 2\sum_{j=1}^n\delta_j+2\delta\leq
4\sum_{j=1}^n\delta_j<\epsilon^2,\eeq

\noidt what is the desired result. \end{proof}

Hence, each orbit of the coadjoint action of \fk U\ going through density matrices with only
finite number of different eigenvalues is a submanifold of \fTs: There is an open \nbhd\ of any
point $\nu$\ of $\mOUr=\mfk U/\mUr$ which coincides with intersection of the embedded \OUr\ into
\fTs\ with an open \nbhd\ of the point $\nu$\ in \fTs.

Another possibility of proving this theorem is indicated in the next Section, where such a proof
for the specific case of \OUr := \PH\ is given.
%%%%%%%%%%%%%%%%%%%%%%%%%%%%%%%%%%%%%%%%%%%%%%%%%%%%%%%%%%%%%%%%
\section{Some other related results}\label{sec;altern}
To give here a proof of the promised closeness of the unitary coadjoint orbit going through any
symmetric trace-class operator of finite rank,\footnote{Compare, however, also Proposition 2.1.5
in \cite{bon}.} we shall use an encoding of the spectral invariants (i.e. the spectra, and their
multiplicities) of these operators into finite positive measures on \bR:

\begin{prop}
The unitary orbits \OUr\ for finite-rank $\rho\in\mfTs$ are closed subsets of \fTs.\zal
\end{prop}

\begin{proof}
 Let us take now
the smooth (although differentiability will not be exploited here) numerical functions
$\rho\mapsto a_n(\rho):=Tr(\rho^{n+2})$ determined for all symmetric trace-class operators
$\rho\in\mfTs$. It is claimed that fixing the infinite sequence $\{a_n(\rho), n=0,1,2,\dots\}$ of
real numbers one can determine the unitary orbit $\mOUr\subset\mfTs$ (on which the numbers $a_n$
are constant: $a_n(\ru\rho \ru^*)\equiv a_n(\rho),\ \forall \ru\in\mfk U,\ \rho\in\mfTs$)
uniquely. This can be seen as follows: The orbit $\mOUr$ for a finite-rank $\rho$ is determined by
the spectral invariants of any $\nu\in\mOUr$, i.e. by its nonzero eigenvalues and their
multiplicities. These might be, however, determined by a measure $\mu_\rho$ on \bR, namely the
(not normalized) measure given by the characteristic function $t(\in\mbR)\mapsto
Tr(\rho^2e^{it\rho})$, the moments of which are exactly the numbers $a_n(\rho)$ . That measure
expressed by the nonzero eigenvalues $\lambda_j$ of $\rho$, and their multiplicities $m_j$, has
the form

\beq \mu_\rho=\sum_j \lambda_j^2\cdot m_j\cdot\delta_{\lambda_j},\eeq

\noidt where $\delta_\lambda$ is the Dirac probabilistic measure concentrated in the point
$\lambda$. It is clear that this measure $\mu_\rho$ determines the orbit uniquely. The uniqueness
of the solution of the Hamburger problem of moments (see \cite[Theorem X.4, and Example 4 in Chap.
X.6]{R&S}) for the moments given by the sequence $\{a_n(\rho), n=0,1,2,\dots\}$ proves, that the
measure $\mu_\rho$ is in turn determined by the sequence $\{a_n(\rho)\}$ uniquely.

Since the functions $\rho\mapsto a_n(\rho)$ are continuous in the trace (and even Hilbert-Schmidt,
and on bounded balls in \fTs\ also in the operator \LH-\ ) topology, the intersection of the
(closed) inverse images $a_n^{-1}[a_n(\rho)]\ (n\in\mbZ_+)$:

\beq {\mOUr} = \bigcap_{n=0}^\infty\{\nu\in{\frak{T}}_s:a_n(\nu)=a_n(\rho)\}\eeq

\noidt is a closed subset of \fTs\ in these (induced) topologies.\end{proof}

Next will be given an independent way of proving the above Theorem \ref{thm;reg-emb}, but only for
a specific case of the orbit \OUr\ with $\rho=P_x$, i.e. for the projective Hilbert space \PH. A
use of that method for other orbits \OUr\ would need calculation of the distance functions
$d_\rho(\ru\rho\ru^*, \rv\rho\rv^*)$ on the riemannian manifolds \OUr\ for a general \rh\ of
finite range.\footnote{Remember that (cf. \EQM) for $\mrh\in\mfTs$ with infinite range the claim
of Theorem \ref{thm;reg-emb} is false!}

\begin{prop}\label{prop;PH}
The unitary orbit \OUr\ going through a one dimensional projection $\mrh:=P_x\ (0\neq x\in\mH)$ is
a submanifold of (i.e. it is regularly embedded into) the space \fTs\ of symmetric trace-class
operators.\zal
\end{prop}
\begin{proof} It is known, that the riemannian distance function on \PH\
is (cf., e.g., the formula (3.2.11) in \EQM): \beq d(P_x,P_y)=\sqrt{2}\arccos\sqrt{Tr(P_xP_y)}.
\eeq

On the other hand, the distance between the same projections in the ``ambient space'' \fTs\ is

\beq\label{eq;dist1} Tr|P_x-P_y|=2[1-Tr(P_xP_y)]^{1/2},\eeq

\noidt what is easily obtained as the sum of absolute values $|\lambda_1|+|\lambda_2|$ of the two
nonzero real eigenvalues (if $P_x\neq P_y$; choose $\lambda_1\geq \lambda_2$ ) of $P_x-P_y$: Since
$Tr(P_x-P_y)=\lambda_1+\lambda_2=0$, one has $\lambda_1=-\lambda_2=:\lambda>0$. Because
$2\lambda^2=Tr[(P_x-P_y)^2]=2[1-Tr(P_xP_y)]$, one obtains $\lambda=\sqrt{1-Tr(P_xP_y)}$, hence the
result \rref dist1~. We see that these two metrics are mutually equivalent.

This implies that the convergence of some sequence $\{P_{y_n}:n\in\mbZ_+\}$ of points of this
orbit to a chosen point $P_x\in\mOUr$ \emph{in the space} \fTs\ means also its convergence {\em on
the orbit} \OUr, what gives the wanted continuity of the inverse $\iota^{-1}$ of the injective
immersion (it was proved earlier in \EQM\ that $\iota$ is an immersion) $\iota: \mfk U/\mfk
U_{P_x}=\mcl O_{P_x}(\mfk{U})\rarw \mPH\subset\mfTs$ (the set \PH\ is taken here in the relative
topology of \fTs). This means, that the injection $\iota$ is a homeomorphism, hence \PH\ is a
submanifold (cf. \cite{bourb;manif}) of \fTs.\end{proof}

It might be useful to formulate an easy generalization of Lemma \ref{lem;proxi}. One can see that
the condition \rref EF~ of ``proximity'' of the two projections $E, F$ was used in the proof of
that lemma for proving the item (i) only. Assuming the conclusion (i), one can formulate a
generalization of Lemma \ref{lem;proxi} valid also for infinite-dimensional projections, and
without any restriction to their mutual ``proximity'':

\begin{prop}\label{prop;EF}
Let $E, F$ be two orthogonal projections in a separable (real, or complex) Hilbert space \H\ with
mutually isomorphic ranges: $E\mH\sim F\mH$. Assume that $E\wedge F=0$, and that for any
one-dimensional projections $P_e\leq E$, and $P_f\leq F$ it is

\beq\label{eq;eF} P_e\cdot F\neq 0,\quad P_f\cdot E\neq 0.\eeq

\noidt Let also the spectrum of $EFE$ be pure-point (i.e. the eigenvectors form a basis of \H).

Then there is an orthonormal decomposition of $E$ to one-dimensional projections $E=\sum_j
P_{e_j}$ (the sum is strongly convergent), to which there is a unique orthogonal one-dimensional
decomposition of $F: \sum_j P_{f_j} = F$ such that

\beq\label{eq;orthog} P_{f_j}P_{e_k}=0\quad {\rm for}\ j\neq k,\qquad P_{f_j}P_{e_j}\neq 0, \eeq
for all values of the indices. \zal\end{prop}

 \begin{proof}
The validity of the proposition in the case of $\dim E=\dim F<\infty$ is seen from the proof of
Lemma \ref{lem;proxi}.  In our case, the proof of the Lemma \ref{lem;proxi} can be essentially
used as a first step for proving our claims also for infinite dimensions. Let $\dim E =\dim F =
\infty.$ The operator $EFE$ restricted to $\mcl E:= E\mH$ has trivial kernel: ${\rm Ker}_{\mcl
E}(EFE)=0$, due to the assumption \rref eF~. Let an orthonormal basis in the subspace $\mcl
E:=E\mH$ consisting of the eigenvectors of $EFE$ be $\{e_j\in\mH: j\in \mathbb N\}$. It exists
because $EFE$ has pure point spectrum. The basis $\{e_j\}$ also determines an orthonormal
decomposition $\{P_{e_j}\}$ of $E$. Then the vectors

\beq\label{eq;f-j} f_j:=\|Fe_j\|^{-1}Fe_j,\qquad \forall\ j\in\mathbb N\eeq

\noidt form an orthonormal system in $\mcl F:=F\mH:$ $\lb f_j|f_k\rb\propto\lb Fe_j|Fe_k\rb= (e_j,
EFE e_k)\ (\forall j,k).$ Let $P_{f_j}$ be the one-dimensional orthogonal projections onto
subspaces of \H\ spanned by $f_j$'s, and define $F_n:=\sum_{j=1}^nP_{f_j}\ (\leq F)$. Let also
$E_n:=\sum_{j=1}^nP_{e_j}\ (\leq E).$

The projections $E_n$ and $F_n$ are both (finite) $n$-dimensional and fulfill the assumptions of
the proposition (by obvious interchange $E\leftrightarrow E_n,\ F\leftrightarrow F_n$). Also it is
$Fe_j=F_ne_j,\ j=1,2,\dots,n$, so that the presently defined $P_{f_j}$'s coincide with those
obtained according to the proof of Lemma \ref{lem;proxi}. It is clear that also the orthogonality
relations \rref orthog~ are fulfilled. It remains to show that

\beq\label{eq;supF} F=s-\lim_{n\rarw\infty}\ \sum_{j=1}^n P_{f_j}:=\bigvee_{n=1}^\infty F_n.\eeq

Obviously, it is $\vee_{n=1}^\infty F_n\leq F$. We have to prove equality in this relation. Assume
that there is a one-dimensional projection $P_f\leq F$ orthogonal to all $F_n: P_f\cdot F_n\equiv
0$. Since, according to \rref eF~, $P_fE\neq 0$, there is at least one $e_k$ contained in the
given orthonormal system $\{e_j\in\mH: j\in \mathbb N\}$ such that $P_fe_k\neq 0$. But
$FP_f=P_fF=P_f$, and any vector $f_k\neq0$ corresponding to $P_{f_k}$ is $f_k\varpropto
Fe_k=F_ne_k\ (\forall n\geq k)$. Consequently, for all $n\geq k$ it is
$P_fF_ne_k=P_fFe_k=P_fe_k\neq0$, what implies $P_f\cdot F_n\neq0\ (n\geq k)$. So that any assumed
$P_f$ orthogonal to all $F_n$'s does not exist, and the equality in \rref supF~ holds.

The uniqueness of $\{P_{f_j}; j\in\mathbb N\}$ corresponding to the decomposition $\{P_{e_j};
j\in\mathbb N\}$ of $E$ and determined by eigenvectors $|e_j\rb$ of $EFE$ in $\mcl E$, with the
stated properties follows from the orthogonality relations \rref orthog~: It is obtained by
orthogonal projecting of the $e_j$'s onto \cl F: $f_j\varpropto Fe_j,\ \forall j$. This proves the
proposition.
 \end{proof}

To see a rather weak connection of the derived properties of considered projections $E, F$ with
their previously discussed mutual ``proximity'', we shall consider an explicit representation of
these projections. It will show also in which way the point-spectrum of the restriction of $EFE$
to $\mcl E:=E\mH$ can be made an arbitrary countable subset of the open interval
$(0,1)\subset\mathbb R$.

\begin{exm}
Let $E$ be an orthogonal projection in a complex Hilbert space \H\ and let $\{e_j:j\in J\}$ (with
an index set $J$ of cardinality $\leq\aleph_0$) be an orthonormal basis in $\mcl E:=\ E\mH$. Let
$E^\bot$ be another orthogonal projection in \H\ with the same ``dimension $J$'' of $\mcl
E^\bot:=\ E^\bot\mH$ and orthogonal to $E:\ E\cdot E^\bot=0$. Let $\{e^\bot_j:j\in J\}$ be an
orthonormal basis of $\mcl E^\bot$. Let us choose an arbitrary set of complex numbers
$\{\alpha_j,\beta_j:j\in J\}$ such that $\alpha_j\cdot\beta_j\neq 0,\ |\alpha_j|^2+|\beta_j|^2=1,\
\forall j$. Let us define in \H\ vectors $f_j:=\alpha_je_j+\beta_je_j^\bot,\ \forall j\in J$. The
vectors $\{f_j:j\in J\}$ form an orthonormal basis in a subspace $\mcl F\subset \mH$ with the
orthogonal projection $F:\ F\mH=\mcl F$. It is clear that the couple of projections $(E;F)$
satisfies the assumptions of the Proposition \ref{prop;EF}, and that the specified sets of vectors
$\{e_j:j\in J\}$ and $\{f_j:j\in J\}$ determine decompositions of $E$, and $F$, respectively,
appearing in the assertions of Proposition \ref{prop;EF}.

Now we see that the spectrum of our positive bounded operator $EFE$ is pure-point and contains the
eigenvalues $\{|\alpha_j|^2: j\in J\}$, with the eigenvectors $e_j\ (j\in J)$. But we could choose
the $\alpha_j$'s arbitrarily with the only restriction $0<|\alpha_j|<1$. Hence, the pure-point
spectrum of $EFE$ with $\dim(E)=\aleph_0$ can be made, in this way, an arbitrary countable subset
of the real interval $(0,1)$.\footnote{Let us note that, according to a theorem by Naimark (cf.
\cite[Chap. 9, Theorem 3.2]{davies}), any positive operator $A: 0\leq A\leq I_{\mcl E}$ defined on
a Hilbert space $\mcl E$ can be extended into the form $A=EFE\rceil_{\mcl E}$, where $E, F$ are
some orthogonal projections in a Hilbert space \H, and $\mcl E=E\mH$.}

To investigate the question of mutual ``proximity'' of projectors $E$ and $F$, let us calculate
first the distance $\|E-F\|_2^2= 2(N-Tr(EF))$ in the case of $|J|=N<\infty$. Due to the
orthogonality relations \rref orthog~, resp. \rref orthog1~, we have (in the Dirac notation)
$Tr(EF)=\sum_{j\in J}|\lb e_j|f_j\rb|^2=\sum_{j\in J}|\alpha_j|^2$. So that, it is:

\beq 0<\|E-F\|^2_2=2(N-\sum_{j\in J}|\alpha_j|^2)<2N, \eeq where every value of the open interval
$(0,2N)$ can be reached without violating our general specification of $(E;F)$. General
projections of the dimension $N$ could reach all values in the closed interval:
$0\leq\|E-F\|^2_2\leq 2N$.

With a help of their chosen representation, we can calculate also the ``proximity'' of the
projections $(E;F)$ in the operator norm, i.e. $\|E-F\|$, what can be used also if $|J|=\aleph_0$.
This can be easily done, because the two-dimensional subspaces spanned by the couples of vectors
$\{e_j;f_j\},\ j\in J$, are all mutually orthogonal. Then the spectrum of $|E-F|$ can be easily
calculated: $|E-F|= \sum_{j\in J}|P_{e_j}-P_{f_j}|$, the spectrum is (cf. proof of the Proposition
\ref{prop;PH}) $\sigma(|E-F|)=\{\sqrt{1-Tr(P_{e_j}P_{f_j})}:j\in J\}$, and the norm is

\beq \|E-F\|=\sup_{j\in J}\|P_{e_j}-P_{f_j}\|=\sup_{j\in J}\sqrt{1-Tr(P_{e_j}P_{f_j})}=\sup_{j\in
J}\sqrt{1-|\alpha_j|^2}. \eeq

\noidt In this case, it is possible to reach all the values $0<\|E-F\|\leq1$ for our projections
(the equality can be reached for $\dim E=\infty$ only). \dovi\end{exm}


\begin{thebibliography}{999}
\bibitem{mars&rati} J. E. Marsden, T. S. Ratiu: {\sl Introduction to
Mechanics and Symmetry}, Springer, New York, 1999.
\bibitem{kiril} A. A.
Kirillov: {\sl Elementy teorii predstavleniyi (Elements of Representations Theory)}, Nauka,
Moscow, 1978, Second edition.
\bibitem{bon}\refer{P. B\'ona:\ {\it Extended Quantum Mechanics}}
{acta phys. slov.}{50}{2000}{1-198} and revised versions in arXiv: math-ph/9909022, resp in
\url{http://sophia.dtp.fmph.uniba.sk/~bona/EQM/eqm8a.pdf}; the original preprint
version is \cite{bon10}.
\bibitem{bon10} P. B\'ona: {\sl Quantum Mechanics with Mean - Field
Back\-grounds}, Preprint No. {\bf Ph10-91}, Comenius University, Faculty of Mathematics and
Physics, Bratislava, October 1991.
\bibitem{Odz&Rat} Anatol Odzijewicz, Tudor S. Ratiu: {\sl Banach Lie-Poisson
 spaces and reduction}, arXiv: math.SG/0210207.
\bibitem{bourb;Lie} N. Bourbaki: {\sl Groupes  et  Alg\`ebres  de  Lie},
Hermann,  Paris, 1972; russ. ed. ``Mir'', Moscow 1976.
\bibitem{bourb;manif} N. Bourbaki: {\sl Varietes differentielles et
analytiques. Fascicu\-le de resultats}, Hermann, Paris, 1967 and 1971; Russ.  ed.  Mir, Moscow,
1975.
\bibitem{jt-schw} J. T. Schwartz: {\sl Nonlinear Functional Analysis},
Gordon and Breach, New York, 1969.
\bibitem{3baby} Y. Choquet-Bruhat,  C. DeWitt-Morette, M. Dillard-Bleick:
{\sl Analysis, Manifolds, and Physics}, Revised edition, North-Holland, Amsterdam - New York -
Oxford, 1982.
\bibitem{R&S} M. Reed, B. Simon: {\sl Methods of Modern Mathematical
Physics}, Vols.I and II, Academic Press, New York - London, 1972 and 1975;
\bibitem{davies} E. B. Davies: {\sl Quantum Theory of  Open  Systems},
Academic,  New York, 1976.
\bibitem{GKMS} J. Grabowski, M. Ku\`s, G. Marmo, T. Shulman: {\sl Geometry of quantum dynamics in
infinite dimension}, arXiv: math-ph/1711.06486.

\end{thebibliography}
\end{document}